\pdfoutput=1
\documentclass[12pt, onecolumn, showpacs, aps, 
nofootinbib, floatfix, prd]{revtex4-2}
\usepackage[a4paper,top=3cm,right=2cm, bottom=3cm,left=2cm]{geometry}
\usepackage{ifpdf}
\usepackage{amsmath,latexsym}
\usepackage{bbm, bm}
\usepackage{amssymb}
\usepackage{graphicx}
\usepackage{xcolor}
\usepackage{tensor}
\usepackage{physics}
\usepackage{comment}
\newcommand{\ave}[1]{\left\langle #1 \right\rangle}
\newcommand{\ensemble}[1]{\left\{ #1 \right\}}
\newcommand{\eqcomma}{\phantom{AA},\phantom{AA}}
\renewcommand{\eqref}[1]{Eq.~(\ref{#1})}
\usepackage[utf8x]{inputenc}
\usepackage[T1]{fontenc}
\usepackage{titlesec}
\usepackage{mathtools}
\usepackage{turnstile}
\usepackage{csquotes}
\usepackage[mathscr]{euscript}
\usepackage[export]{adjustbox}
\newcommand{\nn}{\nonumber}
\usepackage{enumitem}
\begin{document}
\title{Linearized fluctuating hydrodynamics via random polynomials}
\author{Farid Taghinavaz$^a$, Giorgio Torrieri$^b$}
\affiliation{$\phantom{A}^a$ School of Particles and Accelerators, Institute for Research in Fundamental Sciences (IPM),  P.O. Box 19395-5531, Tehran, Iran,\\ 
$\phantom{A}^b$ Universidade Estadual de Campinas (Unicamp), R. Sérgio Buarque de Holanda, 777, Campinas, Brazil, 13083-859.}
\email{$\phantom{A}^a$ ftaghinavaz@ipm.ir,\\
  $\phantom{B}^b$ torrieri@ifi.unicamp.br.}
	\begin{abstract}
We argue that an ensemble of backgrounds best describes hydrodynamic dispersion relations in a medium with few degrees of freedom and is therefore subject to strong thermal fluctuations.  In the linearized regime, dispersion relations become describeable by polynomials with random coefficients.   We give a short review of this topic and perform a numerical study of the distribution of the roots of polynomials whose coefficients are of the order of a Knudsen series but fluctuate in accordance with canonical fluctuations of temperature.    We find that, remarkably, the analytic structure of the poles of fluctuating dispersion relations is very different from deterministic ones, particularly regarding the distribution of imaginary parts with respect to real components.
We argue that this provides evidence that hydrodynamic behavior persists, and is enhanced, by non-perturbative background fluctuations.
	\end{abstract}
	\maketitle
\section{Physical motivation \label{motiv}}
The effect of thermal fluctuations in hydrodynamics is still not understood fundamentally and is of great phenomenological interest due to the seeming applicability of hydrodynamics in small systems \cite{Nagle:2018nvi}, which contradicts naive fluctuation-dissipation thinking, where the mean free path $l_{mfp}$ (a function of viscosity $\eta$ as well as entropy density $s$ and temperature $T$, related to the microscopic relaxation time $\tau_\pi$) is straight-forwardly related to the fluctuation scale $l_{fluct}$ (a function, in the absence of conserved charges, of only the equilibrium temperature $T$ (which, for a relativistic system, also reflects the number of degrees of freedom $N$ per unit volume $V$ via the degeneracy $g$ \cite{ryb,Kovtun:2012rj})
\begin{align}\label{scales}
l_{mfp} \sim \frac{\eta}{sT} \sim \tau_\pi \sim l_{fluct} \sim C_V T^2 \sim \underbrace{\frac{1}{gT}}_{gT^{3} \sim N/V} .
\end{align}
Small system hydrodynamics \cite{Nagle:2018nvi} puts into doubt this $l_{mfp}\sim l_{fluct}$ scaling, especially since emergence of ''collective'' behavior in systems with very few degrees of freedom has also been reported in other situations, such as ultracold atoms \cite{smallhyd1} and even everyday objects (the ''brazil nut effect'' \cite{smallhyd2});  These findings, at very different energy scales, show that perhaps the way we think about collectivity (where first we take the ''many particle limit'' and then the ''small mean free path limit'' \cite{stick,ryb}) needs a rethink.

One way the hierarchy in \eqref{scales} could fail \cite{Dore:2021xqq} is the realization that hydrodynamic degrees of freedom (flow $u_\mu$, heat current $Q_\mu$, temperature $T$, chemical potentials $\mu$ which go into pressure and energy $(p, \varepsilon)$ as well as dissipative tensor $\Pi_{\mu \nu}$) do not match with the observable quantities (energy-momentum tensor current $T_{\mu \nu}$ and conserved current $J_\mu$).   Furthermore, while the latter is specific to each element of the ensemble and extensive, $\beta_\mu=u_\mu/T,\mu$ are considered in the grand canonical limit to be Lagrange multipliers, intensive in the thermodynamic limit.

This suggests that once one applies statistical mechanics limits (ergodicity or each microstate is equally likely \cite{ergodic}) to each cell but allows fluctuations, redundances will appear where each microstate will have a multiplicity of descriptions, not unlike gauge configurations and ghosts in a quantum field \cite{Dore:2021xqq}.    
What an observer interprets as $u_\mu$ in some Landau prescription will be interpreted as heat flow $Q_\mu$ by another observer using the Eckart prescription or $\Pi_{\mu \nu}$ in the MIS prescription.   The important fact is that, since these are multiple descriptions of the same system the dynamics when fluctuations are included, should be invariant with respect to such choices, just like the dynamics of a gauge theory is gauge independent. These have the potential to drastically alter the applicability of hydrodynamics as a function of the number of participants.

While a quantitative understanding of this idea is very involved and elusive, we can make some simple considerations using sound dispersion relations:  Let us think of a sound wave perturbation propagating in a background subject to thermodynamic fluctuations. If $l_{mfp} \gg l_{fluct}$ one can think of the fluctuation as being a localized random perturbation within the sound wave.  We then recover the usual fluctuation-dissipation relations of \cite{Kovtun:2012rj}. But if $l_{mfp} \leq l_{fluct}$ then the background properties (viscosity, speed of sound, etc) fluctuate while the perturbations propagate.  One can then think of the coefficients of that sound mode to be random numbers, whose distribution is given by statistical mechanics.

The consequences of this can be non-trivial:  We know that a pole on the real axis, related to the existence of a real solution, indicates a wave propagating asymptotically.   A pole on the imaginary axis indicates a ''non-hydrodynamic mode'', and a branch cut suggests a violation of (global) univalence, which transits between these regimes.   As we will show, the relation between the distribution of these objects to the distribution of coefficients is rich and non-trivial, indicating that the fluctuating regime is indeed far from a trivial extrapolation of the deterministic one.

To explain these points further, the best interpretation of the dispersion relations is the following Gedanken-experiment:  
let an experimentalist have access to a large ensemble of field configurations of $T_{\mu \nu}(x,t)$ (i.e. many ''events'' where $T_{\mu \nu}(x,t)$  is measured over a fine lattice in ($x,t$).
One can then take the average over a single event (denoted $\ave{...}$.  this can be done by histogramming in bins of $x-x',t-t'$ sampled in a single event) or over the whole ensemble of events (denoted $\ensemble{...}$. This can be done by the usual technique of defining an event specific system of coordinates around the event's center of mass and origin $x_0,t_0$ and building histograms in bins of $x-x_0,t-t_0$ across events ).   In the ensemble average limit (where the Boltzmann equation and molecular chaos apply, or in general where the number of degrees of freedom is effectively infinite) $\ensemble{\ave{...}} \simeq \ave{\ensemble{...}}$ but we will look at deviations from that limit where there is enough statistics to do $\ave{...}$ within a single event.

Thus, let us
measure its two-point function $\ave{T_{\mu \nu}(x,t)T_{\mu \nu}(x',t')}$ event-by-event, take a fast Fourier transformation event-by-event and infer, still event by event, a $\omega = c_n k^n$ to the fast Fourier transform, yielding an ensemble $\ensemble{c_n}$ over all events.  One can do this via a fitting process if the event is large enough or by expressing larger cumulants $\ave{T_{\mu \nu}(x_i,t_i)...T_{\mu \nu}(x_1,t_1)}$ in terms of 2-point functions $\ave{T_{\mu \nu}(x_i,t_i)T_{\mu \nu}(x_{i+1},t_{i+1})}$, in analogy as to what is done with azimuthal angles to measure flow in small systems \cite{cms}.

In the limit where thermal fluctuations are smaller in scale w.r.t. the mean free path, one should obtain the Kubo formula (see next section \ref{seccn}, \eqref{kubo} for $m=1$) where, because of the $\lim_{k\rightarrow 0} k^{-1}...$ limit, only the infrared part of fluctuations contributes to the dynamics (in other words, thermal fluctuations are ''local'' with respect to gradients and the mean free path, as also argued in \cite{functional, ergodic, Dore:2021xqq}).  In the opposite limit, one can think of the perturbation propagating inside a thermal fluctuation.  As we shall argue towards the end of section \ref{seccn} $c_n$s fluctuate independently according to the whole spectral function of $\ave{\tilde{T}_{\mu \nu}(\omega, k)\tilde{T}_{\mu \nu}(\omega', k')}$ (non-hydrodynamic modes and all) \cite{moore}.  
The experimentalist then uses that information, to understand its further dynamics, by extracting the hydrodynamic parameters in each event ($\ensemble{e,p},\ensemble{\eta/s}$ and so on) and using them, as an effective field theory expansion in the Knudsen number to predict the subsequent evolution of an initial perturbation {\em in that given event} via a propagator (which we define in \eqref{jdef}), {\em constructed for that event} (we, of course, assume the experimentalist can also construct perturbations in that event and there are enough DoFs to construct such a perturbation).  

It is immediately evident from the mathematics literature \cite{Nicolaescu2022, Nguyen_2016} how fluctuations affect this conclusion.   As $m \rightarrow \infty$ in \eqref{disp} and in the limit that $\eta/s \sim 1$ it is a classic result \cite{kac} that the roots converge to a circle in the complex plane.   As a result, 
\begin{description}
\item[The fraction] of real roots approaches 0,
\item[The probability] of having {\em at least one root} approaches 1.
\end{description}
This means that in events where \eqref{disp} has real roots for Fourier coefficients corresponding to the initial spacial distribution, that perturbation will propagate asymptotically in the linear regime. In events where the root has an imaginary part, that perturbation will propagate to a scale proportional to the imaginary part (which is large for that perturbation to be ''hydrodynamic'').

In the next section, we will examine more closely what the dispersion relation described in this section looks like, and in the rest of the work, we proceed with a numerical study of it.
\section{Dispersion relations in a fluctuating medium \label{seccn}}
Let us therefore consider a perturbation with a general dispersion relation.   Neglecting chemical potential, dimensionality and the spacetime structure of the dispersion relation forces us into \cite{Grozdanov:2020koi, Grozdanov:2019kge, Grozdanov:2019uhi}
\begin{equation}
\label{disp}
\omega^\pm(k) = -i \sum_{n=1}^m c_{n}^\pm k^n  \eqcomma c_{n}^\pm = e^{\pm \frac{i n \pi}{2}}\order{c_s^{n} \left(  \frac{\eta}{s}\right)^{n-1}}\tau_{micro}^{n-1} \eqcomma \tau_{micro}\sim \frac{1}{T}, 
\end{equation}
where $c_s$ is the speed of sound in that medium. For example, the usual Navier-Stokes dispersion relation is described by
\begin{equation}
c_1=c_s \eqcomma c_2=\order{1}\frac{\eta}{s} \tau.
\end{equation}
We note that we put together Eqs. (7, 8) of \cite{Grozdanov:2020koi} because event by event and with a finite space resolution the ''sound'' and ''diffusion'' modes can only be distinguished on a Bayesian basis.  The ''complex phase'' dependence on $``n"$ is then determined by analyticity \cite{Grozdanov:2019kge}
where $g\sim N$ is the microscopic degeneracy and $N$ is the number of particles. Note that the first real coefficient does not fluctuate in this simplified model because $c_s^{2}=1/3$.

It is well-known that for a system with $N$ particles (for a small system this can be interpreted as the multiplicity), temperature fluctuations are given by (see \cite{landau} and references therein.  Note that, since the multiplcity is observable, we prefer to write the fluctuation in terms of the multiplicity rather than the degeneracy as in Eq. \ref{scales})
\begin{equation}
\label{tfluct}
\ave{(\Delta T)^{2}}=\frac{2 T^{2}}{N c_v},
\end{equation}
where $c_v$ is the heat capacity at constant volume per each degree of freedom. 
 \footnote{
In a theory with a varying number of degrees of freedom, this represents the microscopic degeneracy, e.g. $\sim N_c^2$ for a 't Hooft type model.}

It is therefore natural to assume temperature to fluctuate with a Gaussian or a Poissonian distribution given by a width of the order shown in \eqref{tfluct} and $c_n$ to fluctuate in response 
\begin{equation}
\label{cnfluct}
\ave{(\Delta c_n)^{2}}\sim \frac{\order{c_s^n \left( \frac{\eta}{s} \right)^{n-1}}}{N c_v} \ave{\left( T\right)^{-2 (n-1)}}.
\end{equation}
Using the terminology prevalent in the literature, if one has a current correlator defined by 
\begin{equation}
\label{jdef}
J(\Delta t,\Delta x)= \int d^{3} k  \frac{\exp\left[ i \left( \omega^\pm(k)\Delta t - k \Delta x \right) \right]}{\omega^\pm(k) + i \sum\limits_{n=1}^m c_{n}^\pm k^n},
\end{equation}
a hydrodynamic mode propagates to $\Delta x \rightarrow \infty$ and a non-hydrodynamic mode to a finite $\Delta x$ (with a long-lived non-hydrodynamic mode propagating for $\Delta t \rightarrow \infty$).   The existence of ''non-hydrodynamic'' poles and branch cuts in $J=\ave{T_{ij}(x), T_{ij}(x')}$ has been a topic of recent interest \cite{moore} with recent claims \cite{gavmodes, rochamodes, roymodes, heinzmodes} using a relaxation Boltzmann equation (where fluctuations are not included) pointing to long-lived non-hydrodynamic modes.

Thus, naively, if one neglects fluctuations non-hydrodynamic modes dominate.  However, as $\Delta x \rightarrow \infty$ in \eqref{jdef} isolates real roots a hydrodynamic mode will always emerge, rendering any non-hydrodynamic mode irrelevant.   

Note that according to the ergodic picture of statistical mechanics \cite{ergodic} fluctuation scaling of $c_n$ is independent of the choice of hydrodynamic frame.   Hence, it should not by default be associated with any "model" of hydrodynamics.   The specific transport coefficients of Landau, Eckart, Israel-Stewart, and other models will {\em not} be random fluctuating quantities $c_n$ themselves but rather will be a Bayesian inference of the coefficients $c_n$ and their relationships \cite{Dore:2021xqq}.

This might prove confusing to someone who develops a hydrodynamic EFT from a thermostatic frame and an ''objective'' definition of $u_\mu$ (that is, most of the literature).   This is an important point as it also leads to the ansatz assumed here, where $c_n$s fluctuate {\em independently} rather than follow fluctuations of $T$ and $N$, as expected from the fact that the average values of $c_n$ would be functions of these two parameters.

We note that every $c_n$ in \eqref{disp} comes from a different degree of freedom, something clear from both effective  \cite{ryb} and transport \cite{dnmr} theory.   The dispersion relation whose zeros are represented by \eqref{disp} is directly derivable from an Eigenvalue equation of a characteristic matrix of correlators 
\begin{equation}
\det \left\vert H_{n} \times H_{0m} - k I_{nm}\right\vert = 0, 
\end{equation}
where the term $H_n$ is given by a generalized Kubo formula
\begin{equation}
\label{kubo}
H_{n}\propto \lim_{k \rightarrow 0} \frac{1}{k^n} \frac{d^n}{d k^n} \int e^{ikx} d x \ave{T_{ij}(x)T_{ij}(0)}.
\end{equation}
(in \cite{ryb} these coefficients are denoted by $X_{I_1...I_n}$, in \cite{dnmr} by moments of $f(x,p)$)
as well as an initial gradient $H_{0m}$.   
Given that each derivative will be a highly non-trivial function of microscopic physics and initial conditions, representing these matrix coefficients by random numbers
is a consequence of local equilibrium dynamics being dominated by fluctuation-dissipation relations.  Physically, thermal fluctuations continuously produce sound waves, which interact and deform the background according to the Feynman rules developed. For instance, in \cite{nicolis}
we now ``renormalize the background'' with these sound waves, using methods in \cite{Kovtun:2012rj, stick} and introduce a fluctuation on top of that (the driven fluctuation in \eqref{jdef}).  
Thus the independent random coefficients of the dispersion relation reflect the fact that when the sampling is insufficient to reconstruct the phase space distribution the problem of thermalization can be connected to a random matrix problem \cite{functional}.

In fact, if one thinks of hydrodynamics as a limit of the Boltzmann equation small systems can be thought of as promoting phase space functions $f(x,p)$ to functionals \cite{functional,functional2,functional3} then terms of the dispersion relation of $\order{k^n}$ become related \cite{dnmr} to the probability distribution of the $n-th$ moment of $p$ of the the functional distribution of the perturbation $\delta f(x,p)$, which leads to the sort of random polynomial we examine.

In the conformal case, the absence of intrinsic scales beyond energy/temperature forces the scaling of fluctuations of \eqref{cnfluct} when $\eta/(sT)$ is such that the fluctuation domain is of the order of the sound wave propagation domain. \footnote{Actually, a full model would contain not just the Legendre transform/Lagrange multiplier for energy $T$ but also of 4-momentum, $\beta_\mu=u_\mu/T$. As argued in the introduction and \cite{Dore:2021xqq}, the full fluctuating dynamics will be affected by the redundances implicit in the freedom to reparametrize the spacetime foliation whose structure, analogous to Fadeev-Popov ghosts in Gauge theory, is highly non-trivial.  In this work, we concentrate only on the effect of these redundances on linearized dispersion relations, parametrized by $c_n$. } 

A basic question in that context is, ``Will this disturbance propagate, far in the IR?'' The probability of a disturbance of $T_{\mu \nu}$ to propagate long-distance is therefore represented by the probability of finding a real root in some element of the ensemble, rather than some parameter depending on the choice for the thermostatic frame (something inferred from the ensemble as much as any other observation).

While the study of random polynomials is a developed field of pure mathematics (we give a short overview in the appendix), as far as we know no analytical results exist which are directly relevant to the type of polynomials described in this section.  In this work, we shall proceed numerically. In section 2, we review the  MIS formalism and derive the low-lying sound mode equation which we compare with the random modes solutions provided in section 3. Eventually, we give a comprehensive conclusion with an outlook for future studies. 
\section{Review of MIS  formalism}
By the MIS model, we address an uncharged conformal system characterized by the following constitutive relation
 \begin{align}
     T^{\mu \nu} = \varepsilon u^\mu u^\nu - p \Delta^{\mu \nu} + \Pi^{\mu \nu},
 \end{align}
 where $\Pi^{\mu \nu} = -2 \eta \sigma^{\mu \nu}$ with $\eta$ denoting the shear viscosity and $\sigma^{\mu \nu} = \frac{1}{2} \left(\nabla^\mu u^\nu + \nabla^\nu u^\mu - \frac{2}{3} \Delta^{\mu \nu} \nabla \cdot u\right)$ represents the shear-stress tensor and $\nabla_\mu = \Delta_{\mu \nu} \partial^\nu$. The governing dynamical equations are given by \cite{Grozdanov:2018fic}
 \begin{align}\label{eq: eqs-MIS}
     &\partial_\mu T^{\mu \nu} = 0,\nn\\
     & \tau u^\nu \partial_\nu \Pi^{\mu \nu} + \Pi^{\mu \nu} = - 2 \eta \sigma^{\mu \nu},
 \end{align}
 where $\tau$ is the relaxation time associated with $\Pi^{\mu \nu}$ approaching its on-shell value, $-2\eta\sigma^{\mu \nu}$.  $\Pi^{\mu \nu}$ is a regulator field with the lifetime $\tau$, being the first non-conserved operator in the non-hydrodynamic spectrum.  Our analysis focuses on the sound channel in which by solving the  \eqref{eq: eqs-MIS} for small perturbations, the equation takes the following form \cite{Grozdanov:2018fic}
 \begin{align}\label{eq: sound-MIS}
      \omega^3 + \frac{i \omega^2}{\tau} - \left(c_s^2 + \frac{\gamma_s}{w \tau}\right) \omega k_z^2 - \frac{i c_s^2 k_z^2}{\tau}= 0,
 \end{align}
 where $c_s^2 = \partial \varepsilon/\partial p$ is the speed of sound, $\gamma_s = 4\eta/3$ and $w = \varepsilon + p$ is enthalpy.  We can make the substitutions
 \begin{align}
  \omega = i \frac{\beta}{\tau}, \qquad k_z^2 \equiv z = \frac{\tilde{z}}{c_s^2 \tau^2}, \qquad X \equiv -1 + \frac{\gamma_s}{8 c_s^2 \tau w},   
 \end{align}
resulting in the following dispersion relation
\begin{align}\label{eq: disp-sound-MIS}
    \beta^3 + \beta^2 +\tilde{z} \, \beta (9+8X) + \tilde{z} = 0.
\end{align}
In what follows, as a part of the computation, we want to compare the random modes structure with the infrared (IR) solutions of the \eqref{eq: disp-sound-MIS}.
\section{Algorithm to find the probability of  propagating modes}
In our calculations, we utilize the assumption of conformal symmetry, which implies that $\partial p /\partial \varepsilon = c_s^2 = 1/3$. Furthermore, we scale the frequency $``\omega"$ by the temperature, denoted as $``T"$. We aim to calculate the probability of propagation for real modes within the infrared (IR) limit of the MIS theory. We perform this analysis for two specific sets of parameters: i) $\tau_1 T = (2-\ln2)/(2\pi)$, which corresponds to the $\mathcal{N}=4$ result \cite{Baier:2007ix}, and ii) $\tau_2 T= 0.1/(2\pi)$. We investigate the behavior at both high and low momenta for each parameter set. We examine each case using series expansions of order 4 and order 10 in momentum to assess the influence of series order on our findings. Here, we outline the steps followed in our analysis.
\begin{enumerate}
    \item We generate the modes using the following series \cite{Grozdanov:2020koi, Grozdanov:2019kge, Grozdanov:2019uhi}
    \begin{align}\label{eq: random-omega}
        \omega_s(k) = -i \sum\limits_{n=1}^s c_n e^{i\frac{n \pi}{2}}k^n - i \sum\limits_{n=1}^{\frac{s}{2}} d_{2n} k^{2n}.
    \end{align}
The choice of this series is due to the mathematical properties of dispersion relation at low momenta which we assume to be free of any singular characters like branch-point, branch-cut, etc. Indeed, the series in \eqref{eq: random-omega} generates the sound and the diffusive channel because in an experimental setup, we can not separate these channels.  Based on the dimensionality and space-time structure of each mode, the $c_n$ and $d_n$ coefficients must take the following form
\begin{align}\label{eq: scaling-cn}
    (c_n, d_n) =  (a_n, b_n) \times c_s^n (\frac{\eta}{s})^{n-1} \tau^{n-1}.
\end{align}
The coefficients $(a_i, b_i$) are fluctuating numbers corresponding to the first (second) term in the R.H.S. of \eqref{eq: random-omega}. These fluctuations are essentially thermal fluctuations because they are dimensionful quantities. In a conformal theory (which we take here) with only one scale, such as temperature, any dimensionful fluctuation can be converted into thermal fluctuation. 

\item For our later purpose, we need the 4th and 10th expansion of \eqref{eq: random-omega} . In 4th order, this series takes the following form
    \begin{align}\label{eq: random-function-o4}
        \omega_{4}(k) = a_1 c_s k + i (a_2 -b_2)\frac{\eta}{s} \tau c_s^2  k^2 - a_3  (\frac{\eta}{s})^2 \tau^2 c_s^3 k^3 - i (a_4 + b_4)  (\frac{\eta}{s})^3 \tau^3 c_s^4 k^4 + \mathcal{O}(k^5).
    \end{align}
For order 10, it becomes
\begin{align}\label{eq: random-function-o10}
        \omega_{10}(k) &= a_1 c_s k + i (a_2 -b_2)\frac{\eta}{s} \tau c_s^2  k^2 - a_3  (\frac{\eta}{s})^2 \tau^2 c_s^3 k^3 - i (a_4 + b_4)  (\frac{\eta}{s})^3 \tau^3 c_s^4 k^4 + a_5 (\frac{\eta}{s})^4 \tau^4 c_s^5  k^5 \nn\\
        &+ i (a_6 -b_6)(\frac{\eta}{s})^5 \tau^5 c_s^6  k^6 - a_7  (\frac{\eta}{s})^6 \tau^6 c_s^7 k^7 - i (a_8 + b_8)  (\frac{\eta}{s})^7 \tau^7 c_s^8 k^8 + a_9 (\frac{\eta}{s})^8 \tau^8 c_s^9  k^9\nn\\
        & + i (a_{10} -b_{10})(\frac{\eta}{s})^9 \tau^9 c_s^{10}  k^{10}.
    \end{align}
It is worth mentioning that each term in the series mentioned above fluctuates independently, as they originate from the corresponding n-point functions of $T_{\mu \nu}$. We take $\eta/s = 1/(4\pi)$ in our numerical setup.

 \item We employ a discretization of $k = 4 \pi T m/1000$ for our analysis. For low-momentum (LM) results, the algorithm is applied to $m = [1, 2, \cdots, 100]$, and for high-momentum (HM) it runs for $m = [300, 301, \cdots, 500]$. We take each momentum bin is independent of its neighbors, similar to a random walk.
 
 \item The only parameter characterizing the background is temperature (no conserved charges) whose extensive lagrange multiplier is the temperature.
 The variation of other thermodynamic quantities therefore depends on the variation of temperature
\cite{landau2013statistical} as well as the number of degrees of freedom of the system $N$. The temperature distribution follows a Gaussian distribution with the deviation provided in \eqref{tfluct}. We assume $c_v = g$, where $g$ represents the degrees of freedom. We set $g=6$, accounting for three momentum directions and three for spatial displacements. Given that all terms are proportional to $T$ and are positive, we can adopt the following distribution for random and dimensionless $(a_n, b_n)$ numbers
\begin{align}\label{eq: dist-T-ab}
    \mathcal{D}(N, x) = \frac{1}{2} \sqrt{\frac{g N}{\pi}} \frac{e^{-\frac{g N}{4} (x-1)^2}}{1 - \frac{\text{Erf}(\frac{\sqrt{g N}}{2})}{2}}, \qquad x \geq 0,
\end{align}
where $N$ counts the number of particles, and $\text{Erf}(x) = \frac{2}{\sqrt{\pi}} \int\limits_0^x \, e^{-t^2} dt$ is the error function used to normalize the distribution.

\item In each momentum bin, we generate $10^5$ results, with the random coefficients selected according to the distribution \eqref{eq: dist-T-ab}. After generating the series as given in \eqref{eq: random-function-o4} or \eqref{eq: random-function-o10}, we obtain their real and imaginary parts, enabling us to analyze them in various ways.

\item  After obtaining the distribution of roots in each momentum bin, we fit the distribution of either the real or imaginary parts to a Gaussian distribution of the form:
\begin{align}\label{eq: fit-Gaussian}
    \mathcal{D}_G (x_{0_{m}}, \sigma_m; x) = \frac{1}{\sqrt{2 \pi \sigma_m^2}} e^{-\frac{(x - x_{0_{m}})^2}{2 \sigma_m^2}},
\end{align}
to get the optimal mean $x_{0_{m}}$ and standard deviation $\sigma_m^2$. The quality of the propagating modes is evaluated by the ratio $x^{\text{Im}}_{0_{m}}/x^{\text{Re}}_{0_{m}}$. Modes with a smaller ratio deserve to be better propagating modes.

\item Some parts of our results are devoted to comparing the generated series with hydro solutions or the low momentum expansion of solutions of the \eqref{eq: disp-sound-MIS}. For $\eta/s = 1/(4\pi)$ and $c_s^2 = 1/3$, the order 4 series become 
    \begin{align}\label{eq: mode-hydro-MIS}
        &\Omega_{1, 2}^{(4)}(k) = \pm \bigg(\frac{k}{\sqrt{3}} + \frac{k^3}{24 \sqrt{3} \pi^2 T^2} \left( -1 + 4 \pi T \tau \right) \bigg) - \frac{i k^2 }{6 \pi T}  - \frac{i k^4 \tau }{18 \pi^2 T^2} \left(  1 - \pi T \tau\right),\nonumber\\
        &\Omega_3^{(4)}(k) = - \frac{i}{\tau} + \frac{i k^2}{3 \pi T} - \frac{ i k^4 \tau }{9 \pi^2 T^2} \left( - 1 +  \pi T \tau\right).
    \end{align}
    Here, $\Omega_3^{(4)}(k)$ is a purely imaginary, damped mode, while $\Omega_{1, 2}^{(4)}(k)$ modes are propagating modes with some attenuation. For order 10 expansion, the hydro series takes the following form
    \\
    \begin{align}\label{eq: hydro-MIS-o10}
 \Omega_{1, 2}^{(10)}(k) = \pm \bigg(&\frac{k}{\sqrt{3}} + \frac{k^3}{24 \sqrt{3} \pi^2 T^2} \left( -1 + 4 \pi T \tau \right)-\frac{k^5 \left(8 \pi  \tau  T (2 \pi  \tau  T (4 \pi  \tau  T-9)+3)+1\right)}{1152 \sqrt{3} \pi^4 T^4}\nonumber\\
  & +\frac{k^7 \bigg(4 \pi  \tau  T (4 \pi  \tau  T (8 \pi  \tau  T (\pi  \tau  T-5) (4 \pi  \tau  T-5)-25)-5)-1\bigg)}{27648 \sqrt{3} \pi^6 T^6} \nonumber\\
   &\hspace{-2.25cm}-\frac{k^9 \bigg(16 \pi  \tau  T \left(2 \pi  \tau  T (4 \pi  \tau  T (2 \pi  \tau  T (8 \pi  \tau  T (2 \pi  \tau  T-7) (4 \pi  \tau  T-35)-1225)+245)+49)+7\right)+5\bigg)}{2654208 \sqrt{3} \pi^8 T^8}\bigg), \nonumber\\
   & - \frac{i k^2 }{6 \pi T}  - \frac{i k^4 \tau }{18 \pi^2 T^2} \left(  1 - \pi T \tau\right) -\frac{i k^6 \tau ^2 (\pi  \tau  T (\pi  \tau  T-4)+2)}{54 \pi ^3 T^3}  \nn\\
  &  + \frac{ i k^8 \tau ^3 \left(\pi  \tau  T (\pi  \tau  T (\pi  \tau  T-9)+15)-5\right)}{162 \pi ^4 T^4}\nn\\
  &-\frac{i k^{10}\tau ^4 \left(\pi  \tau  T (\pi  \tau  T-2) (\pi  \tau  T (\pi  \tau  T-14)+28)+14\right)}{486 \pi ^5 T^5},\nn\\
  \Omega_3^{(10)}(k) = - \frac{i}{\tau} & + \frac{i k^2}{3 \pi T} - \frac{ i k^4 \tau }{9 \pi^2 T^2} \left( - 1 +  \pi T \tau\right) + \frac{i k^6 \tau ^2 (\pi  \tau  T (\pi  \tau  T-4)+2)}{27 \pi ^3 T^3}\nn\\
  & + \frac{i k^8 \tau ^3 \left(5-\pi  \tau  T (\pi  \tau  T (\pi  \tau  T-9)+15)\right)}{81 \pi ^4 T^4} \nn\\
  &+ \frac{i k^{10}\tau ^4 \left(\pi  \tau  T (\pi  \tau  T-2) (\pi  \tau  T (\pi  \tau  T-14)+28)+14\right)}{243 \pi ^5 T^5}.
\end{align}
\end{enumerate}
    
\begin{figure}
    \centering
    \includegraphics[width=0.475\textwidth]{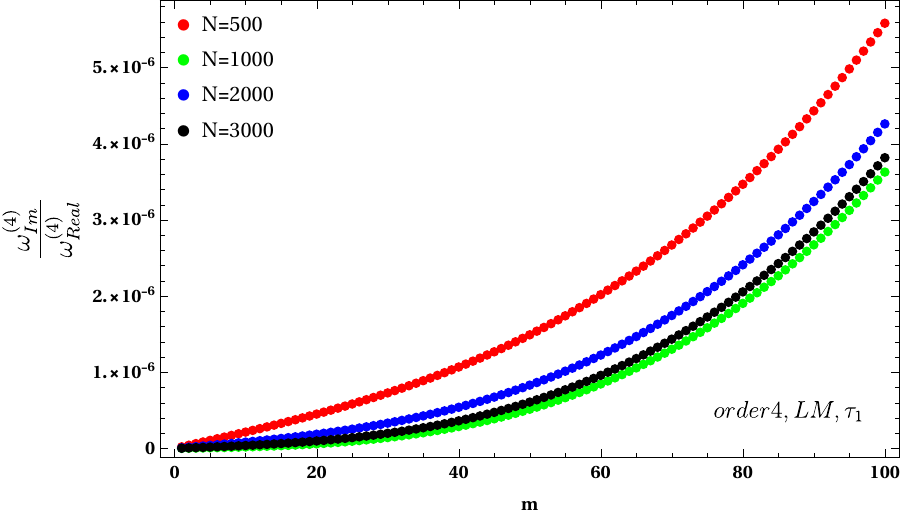}
\hspace{0.3cm}
    \includegraphics[width=0.475\textwidth]{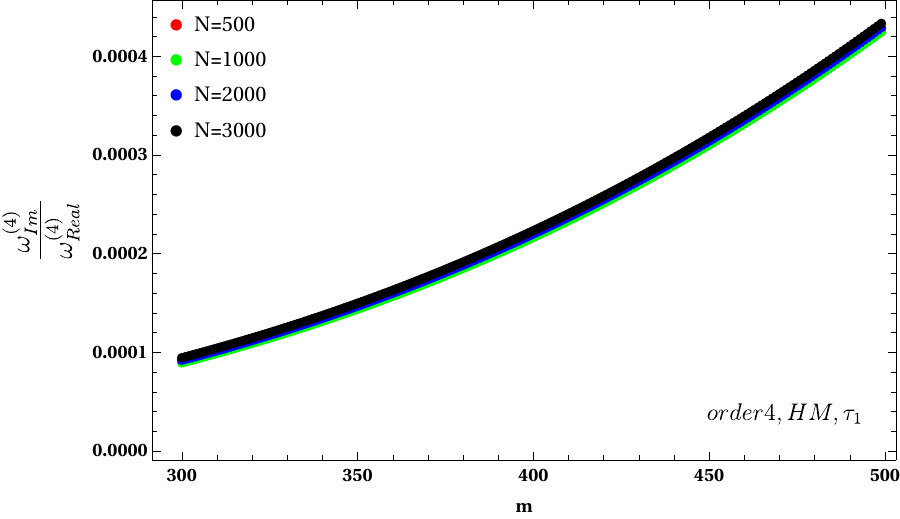}\\
    \vspace{0.5cm}
    \includegraphics[width=0.475\textwidth]{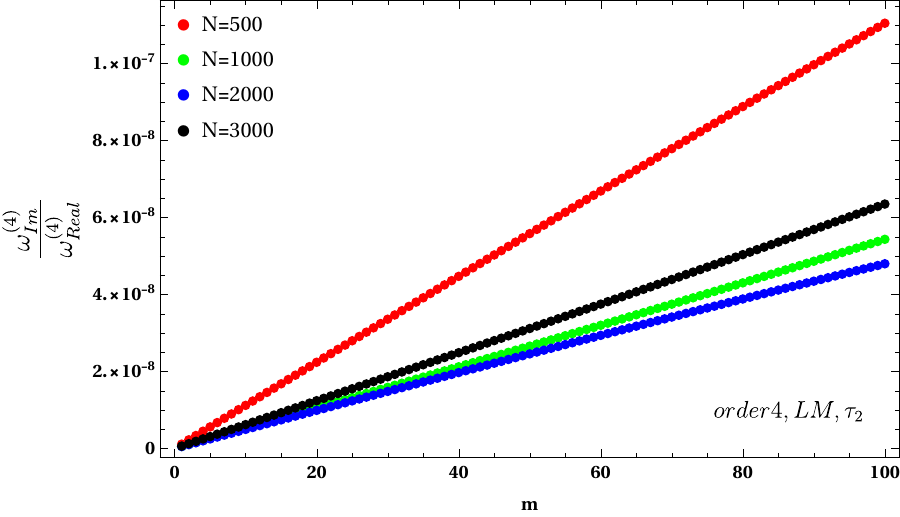}
\hspace{0.3cm}
\includegraphics[width=0.475\textwidth]{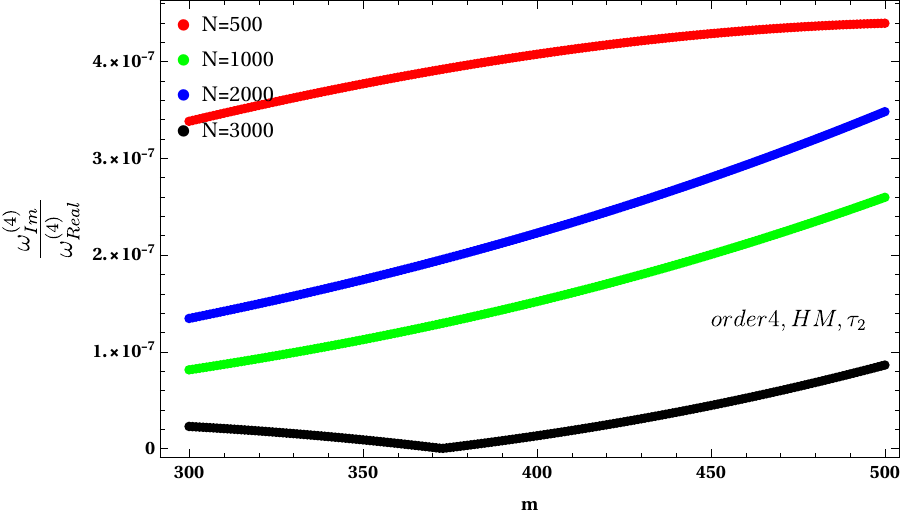}
\caption{The ratio $\omega_{\text{Im}}^{(4)}/\omega_{\text{Real}}^{(4)}$ for various particle numbers in the 4th order series expansion. The top plots display the results for $\tau_1$, and the bottom plots show the results for $\tau_2$. The left column corresponds to low-momentum (LM) bins, and the right column represents high-momentum (HM) bins.}
    \label{fig: plot-im-re}
\end{figure}

After setting all the stages, we are now exploring different cases. In Figure \ref{fig: plot-im-re}, we illustrate the average ratio $\omega_{\text{Im}}^{(4)}/\omega_{\text{Real}}^{(4)}$ for various particle numbers in both low-momentum (LM) and high-momentum (HM) bins, with $\tau_1 = (2-\ln{2})/(2\pi T)$ and $\tau_2 = 0.1/(2\pi T)$. The top panel corresponds to $\tau_1$, while the bottom panel represents $\tau_2$. Also, the left column displays the LM results, and the right column shows the HM results. To calculate this ratio, we generated roots based on \eqref{eq: random-function-o4} for LM bins with $m = (1, \cdots, 100)$ and HM bins with $m = (300, \cdots, 500)$. We then extracted the imaginary and real parts and fitted them to the optimal Gaussian distribution as shown in \eqref{eq: fit-Gaussian}. Afterward, we divided the mean values of each part and organized them according to the value of ``$m$'' for different particle numbers. The results indicate a mild increase with respect to momentum, while the number of particles has a negligible impact. For $\tau_1$, there is a significant difference between the low-momentum (LM) and high-momentum (HM) results. In contrast, for $\tau_2$, the results for both LM and HM fall within the same range. We shall elaborate on this fact in the next few lines.

\begin{figure}
    \centering
    \includegraphics[valign =t, width=0.48\textwidth]{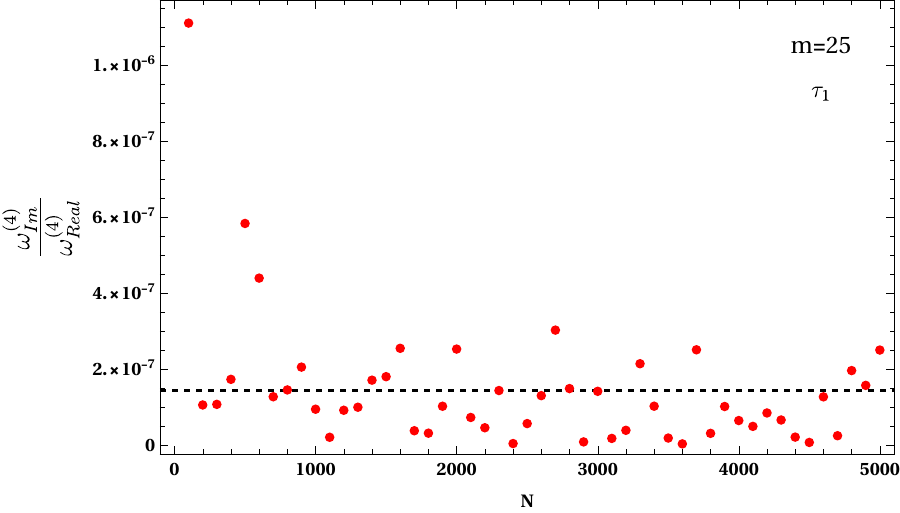}
    \hspace{0.4cm}
    \includegraphics[valign = t, width=0.48\textwidth]{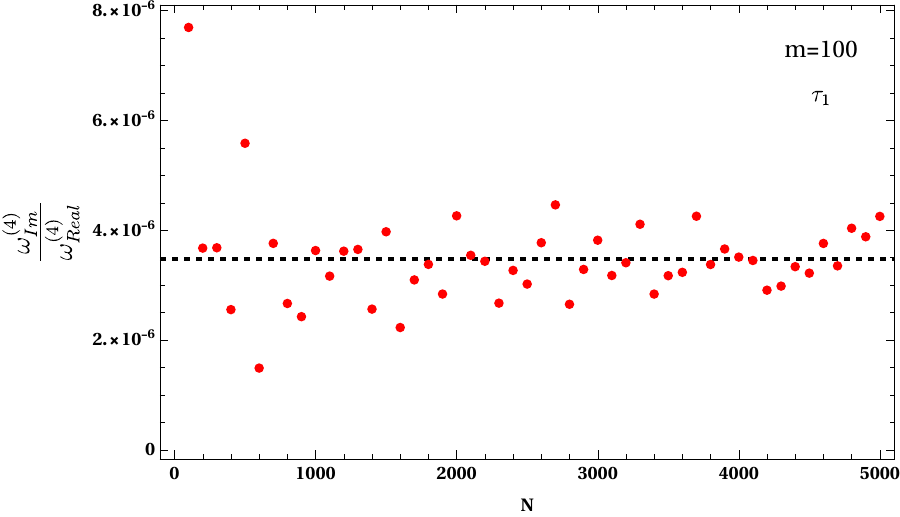}\\
    \vspace{0.5cm}
    \includegraphics[valign =t, width=0.48\textwidth]{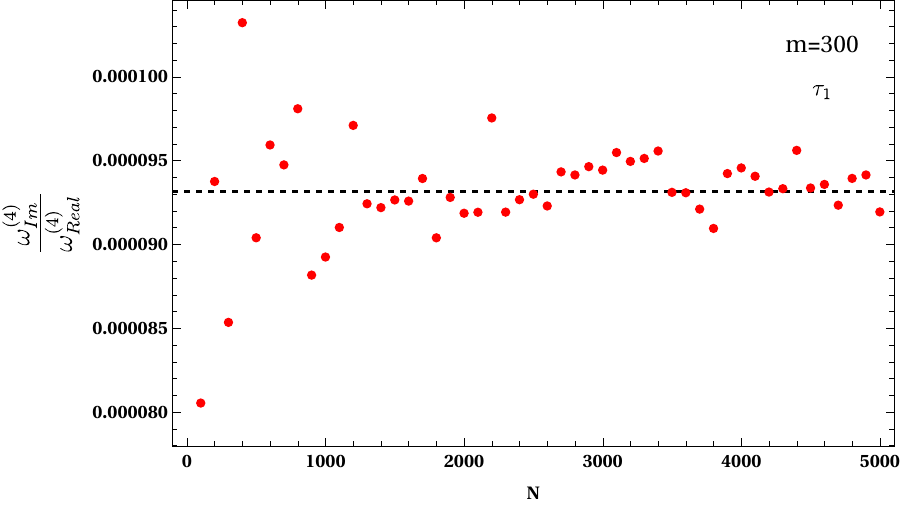}
\hspace{0.4cm}
    \includegraphics[valign =t, width=0.48\textwidth]{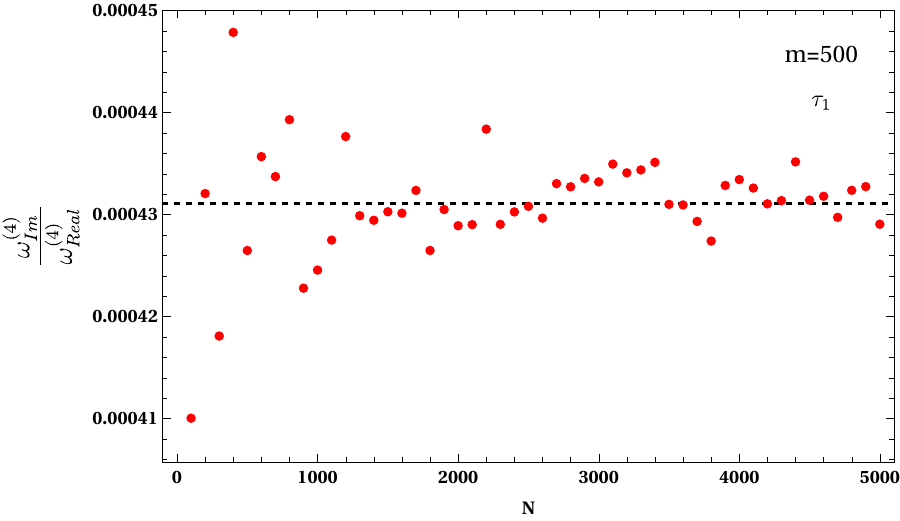}
    \caption{ Plot of $\omega^{(4)}_{\text{Im}}/\omega^{(4)}_{\text{Real}}$ in terms of particle numbers for specific momentum bins. The results are shown for the $\tau_1$ case in the 4th order expansion, utilizing the \eqref{eq: random-function-o4}.}
    \label{fig: plot-im-re-vs-N}
\end{figure}

From Figure \ref{fig: plot-im-re}, we observe an oscillatory pattern in the ratio $\omega_{\text{Im}}^{(4)}/\omega_{\text{Real}}^{(4)}$ as a function of $N$. In Figure \ref{fig: plot-im-re-vs-N}, we present this ratio across the full range of particle numbers for specific momentum bins at $\tau_1$. The horizontal dashed line represents the average value of this ratio. The results of $\tau_2$ behave similarly.

\begin{figure}
    \centering
\includegraphics[width=0.475\textwidth, valign =t]{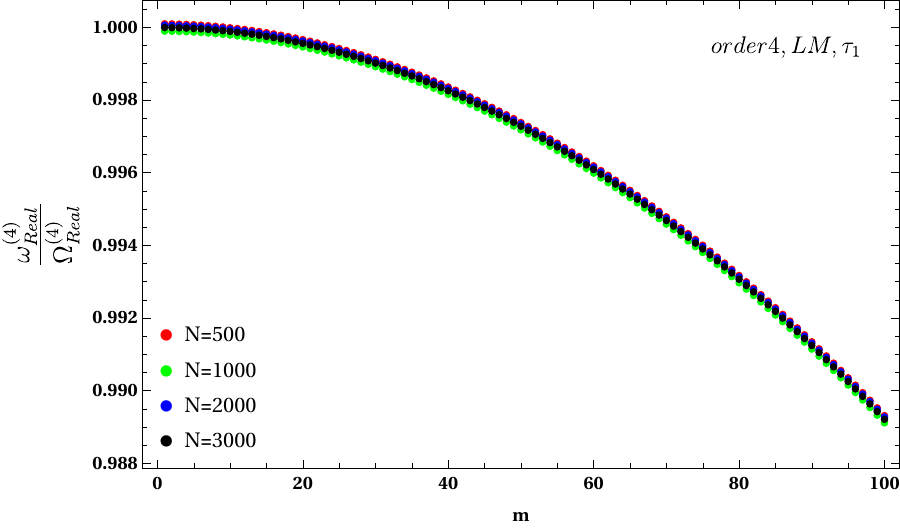}  
    \hspace{0.4cm}
    \includegraphics[width=0.48\textwidth, valign =t]{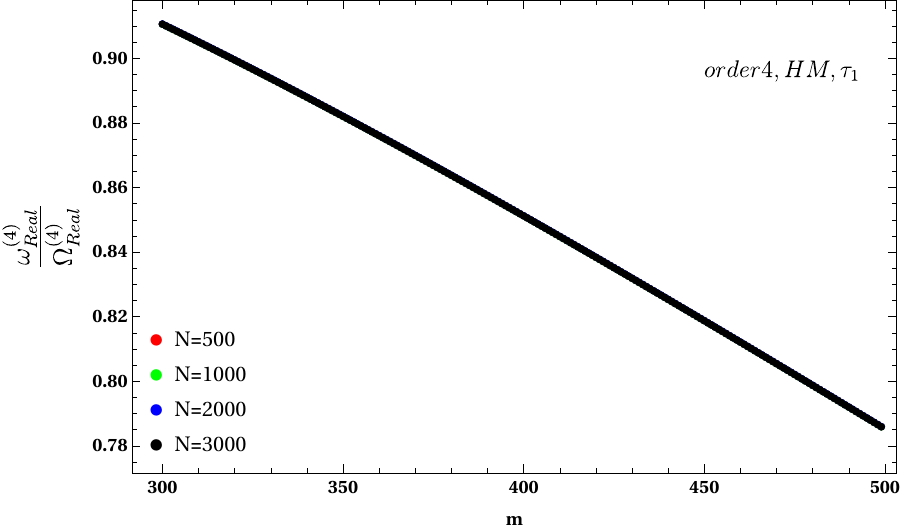}\\
    \vspace{0.6cm}
\includegraphics[width=0.475\textwidth, valign =t]{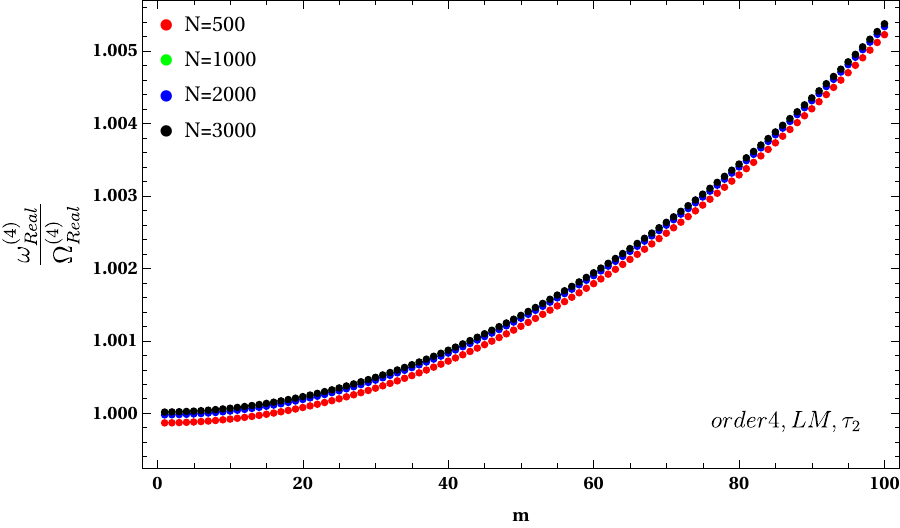}  
    \hspace{0.4cm}
    \includegraphics[width=0.48\textwidth, valign =t]{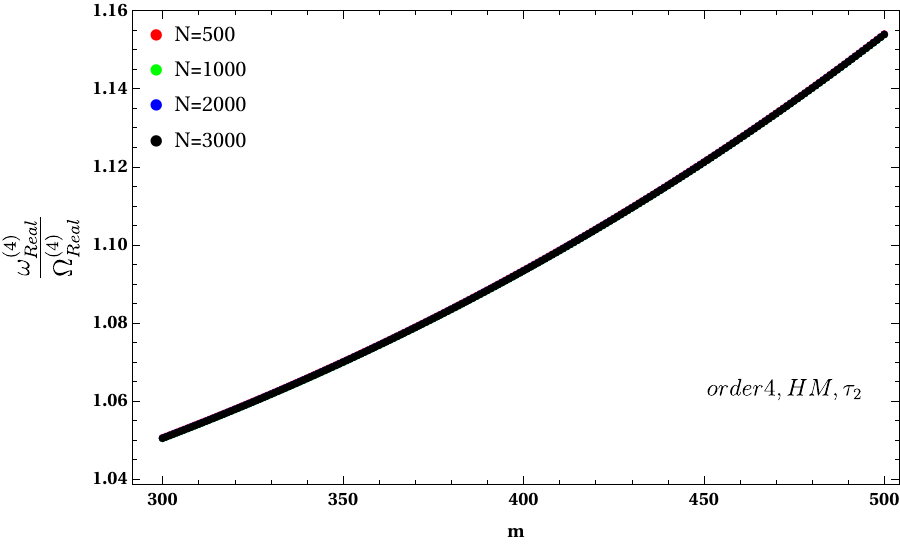}\\
    \vspace{0.6cm}
    \includegraphics[width=0.48\textwidth, valign =t]{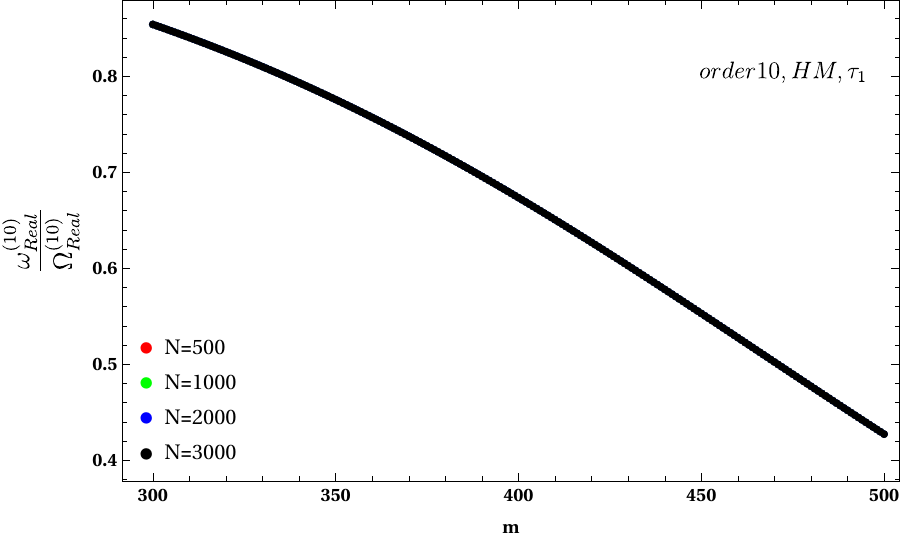}
    \caption{The ratio $\omega^{(4)}_{\text{Real}}/\Omega^{(4)}_\text{Real}$ for various particle numbers. The top row and the very bottom plot correspond to $\tau_1$ and the middle plots belong to the order 4 expansion calculated with $\tau_2$.}
    \label{fig: plot-hydro-real-vs-m}
\end{figure}

In Figure \ref{fig: plot-hydro-real-vs-m}, we examine the ratio between the real parts of the randomly generated modes, as described in \eqref{eq: random-function-o4}, and their counterparts in the hydrodynamic modes, presented in \eqref{eq: mode-hydro-MIS}, for various particle numbers in either LM or HM bins. Each plot represents a specific case of $\tau$ with the HM or LM bin, as indicated. In the top row for $\tau_1$, we see that as momentum increases, the ratio $\omega^{(4)}_{\text{Real}}/\Omega^{(4)}_\text{Real}$ decreases and even higher-order expansions yield smaller values, as shown in the very bottom plot. The results for $\tau_2$ in the middle row demonstrate that going to HM yields larger values, and the expansion order 10 gives no sizable difference with the order 4. For $\tau_1$, this ratio is less than one, while for $\tau_2$, the ratio is greater than one.  Comparing the very top right panel with the very bottom one in Figure \ref{fig: plot-hydro-real-vs-m}, has shown that at HM our ability to predict propagating modes diminishes because different orders produce different results for $\tau_1$. Its message is that we can no longer rely on ansatz \eqref{eq: random-omega} in this region. This is because, intuitively, at high momenta, we pass the radius of the convergence of the sound hydro series which invalidates the use of ansatz \eqref{eq: random-omega}. More interestingly, for the MIS model, the regions of convergence and analyticity are identical \cite{Heydari:2024qzc}, and moving beyond convergence means entering a non-analytic region.  This fact doesn't matter to thermal fluctuations and only the momentum running pushes us into the non-analytic, divergence zone. The size of this region depends on $\tau/T$. For instance, when $\tau = \tau_1$ the points beyond $m \gtrapprox 300$ lie within the non-analytic zone, whereas for $\tau = \tau_2$ the points beyond $m \gtrapprox 1000$ are inside the non-analytic zone. That's why we don't include the $``\mbox{order 10}, HM, \tau_2"$ part in Figure \ref{fig: plot-hydro-real-vs-m}.

\begin{figure}
    \centering
\includegraphics[width=0.475\textwidth, valign =t]{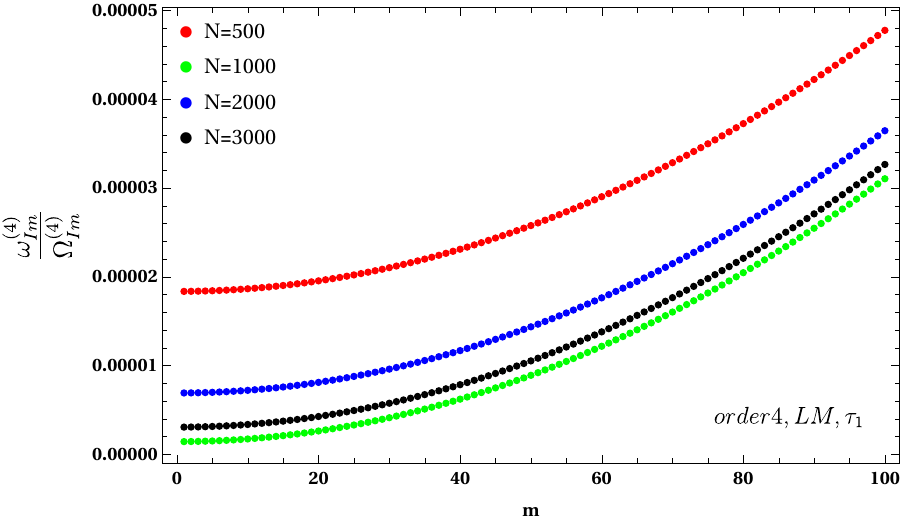}  
    \hspace{0.4cm}
    \includegraphics[width=0.48\textwidth, valign =t]{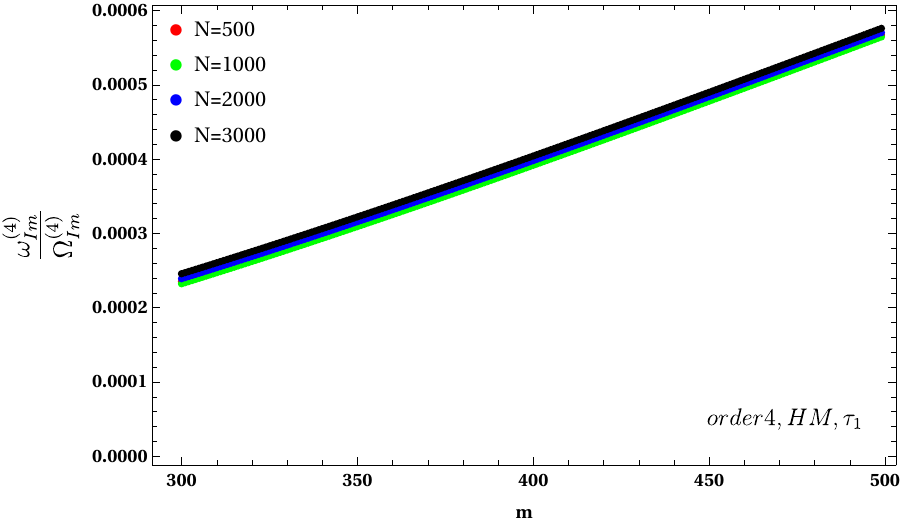}\\
    \vspace{0.5cm}
\includegraphics[width=0.475\textwidth, valign =t]{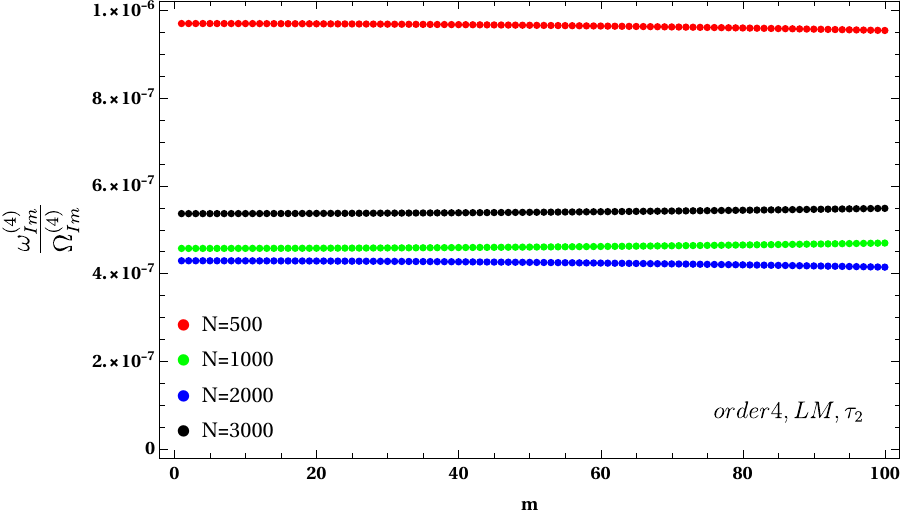}  
    \hspace{0.4cm}
    \includegraphics[width=0.48\textwidth, valign =t]{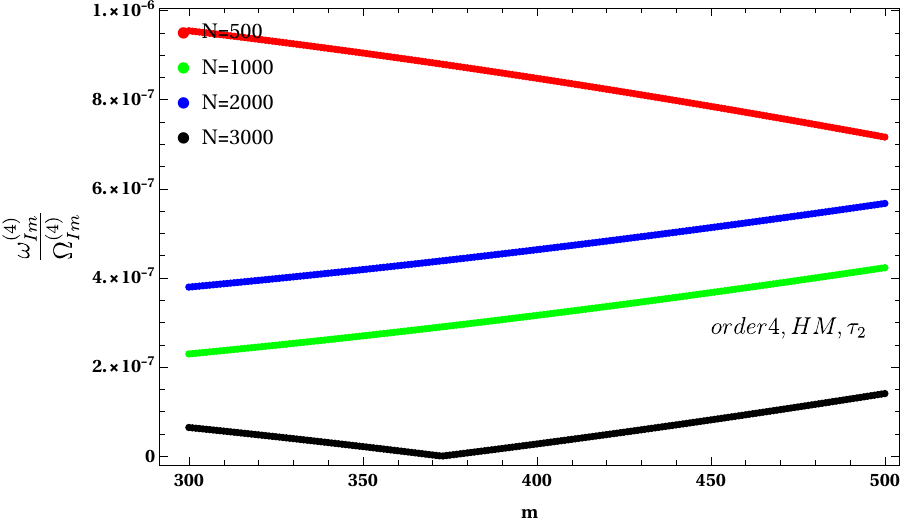}\\
    \vspace{0.5cm}
    \includegraphics[width=0.48\textwidth, valign =t]{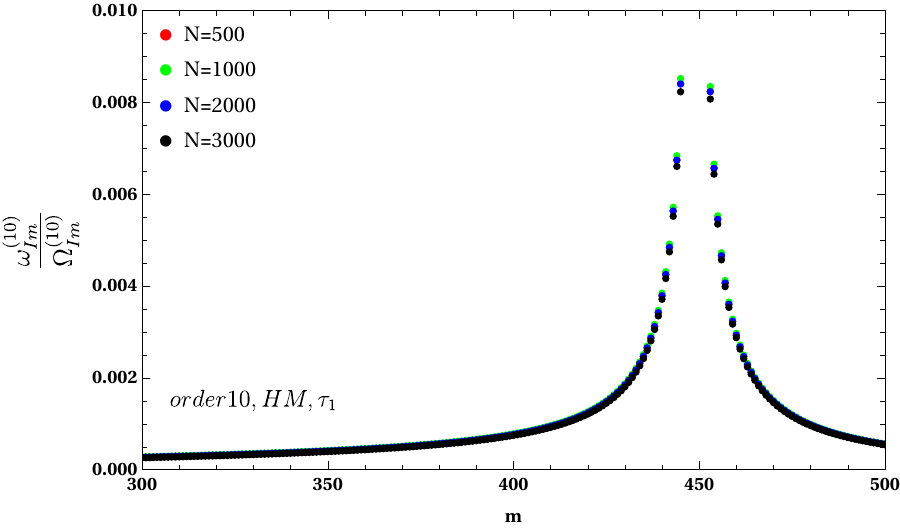}
    \caption{The ratio $\omega^{(4)}_{\text{Im}}/\Omega^{(4)}_\text{Im}$ for various particle numbers. The top row and the very bottom plot correspond to $\tau_1$ and the middle plots belong to the order 4 expansion calculated with $\tau_2$.}
    \label{fig: plot-hydro-im-vs-m}
\end{figure}

In Figure \ref{fig: plot-hydro-im-vs-m}, we display the findings for the imaginary components. In comparison to the real components, the values are considerably smaller. At $\tau_1$, the results exhibit a greater sensitivity to the momentum bin, whereas this sensitivity is smaller in the $\tau_2$ plots. Furthermore, for the 10th-order expansion at $\tau_1$ in the HM bin, there is a noticeable bump around a certain value, which arises from the zero value of $\Omega^{(10)}_{\text{Im}}$.

\begin{figure}
    \centering
\includegraphics[width=0.475\textwidth, valign =t]{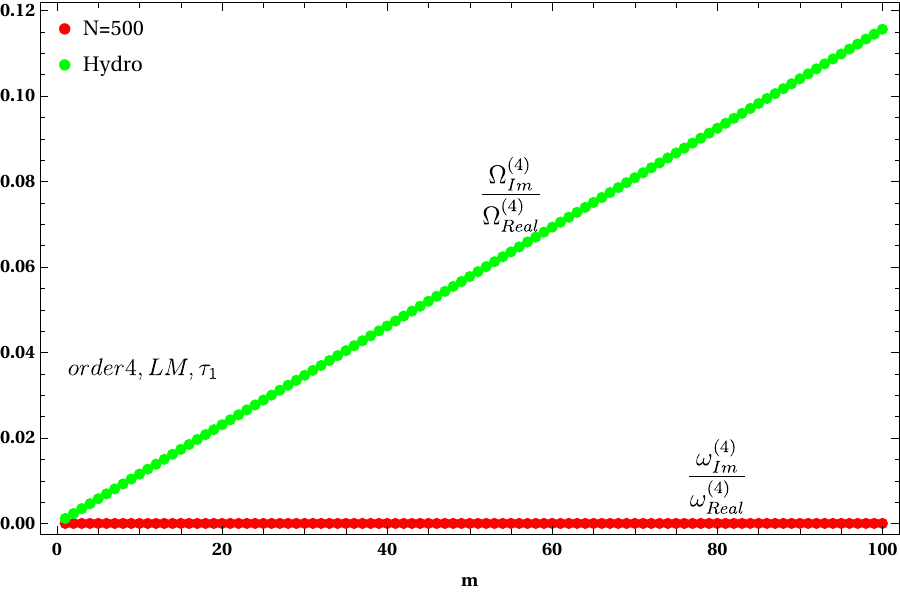}  
    \hspace{0.3cm}
\includegraphics[width=0.475\textwidth, valign =t]{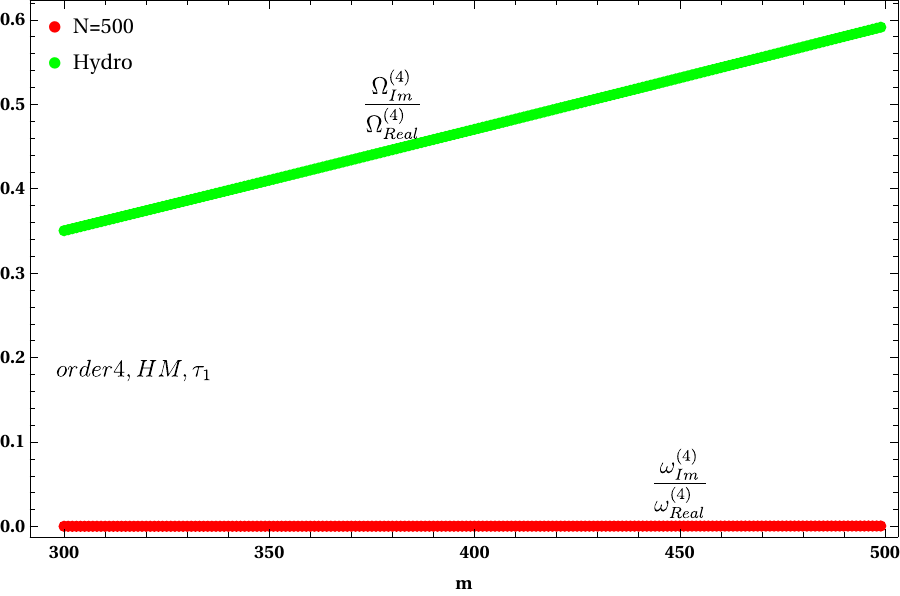}\\
\vspace{0.5cm}
\includegraphics[width=0.475\textwidth, valign =t]{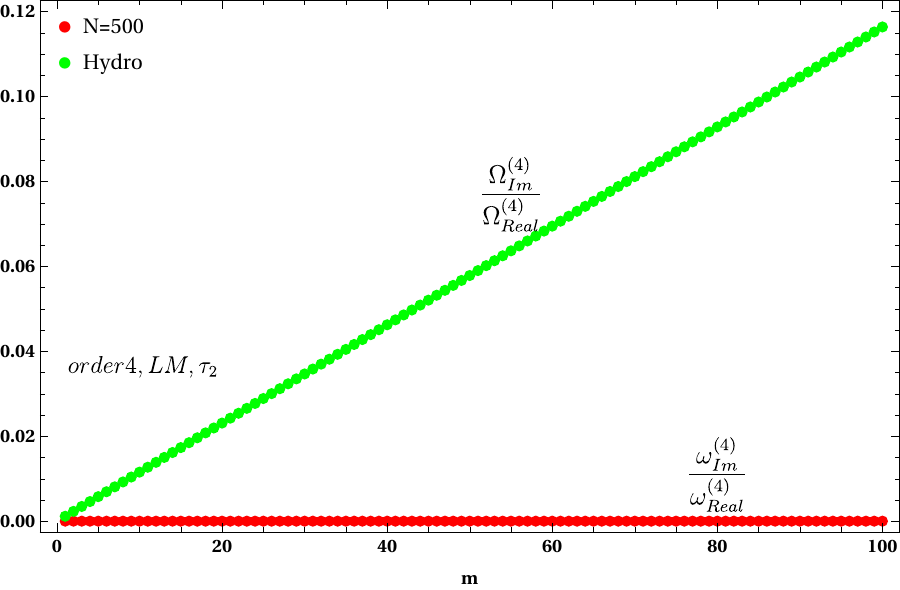}  
    \hspace{0.3cm}
\includegraphics[width=0.475\textwidth, valign =t]{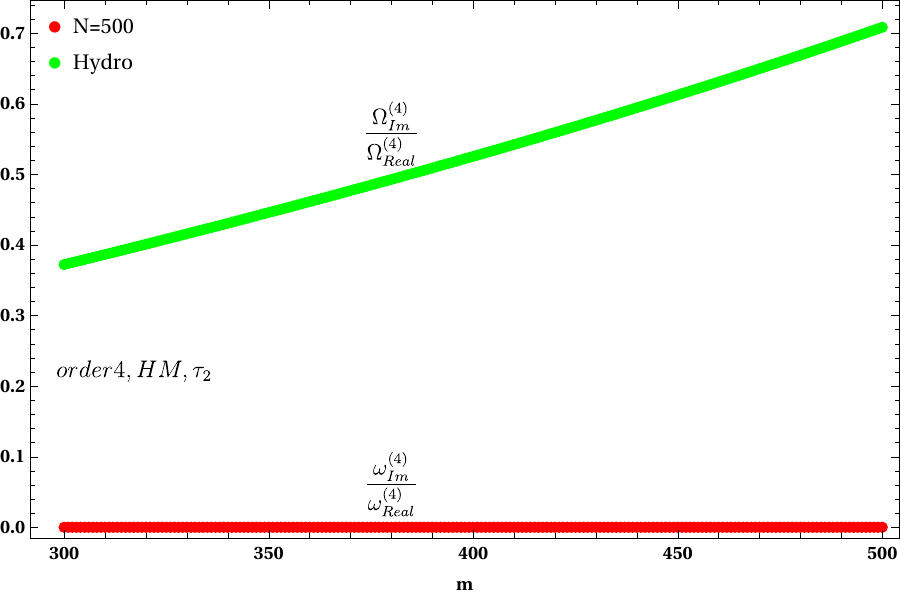}
    \caption{Comparing the imaginary to real parts for the hydro modes and the randomly generated modes. The top (bottom) row corresponds to $\tau_1$ ($\tau_2$). The left (right) panels address the LM (HM) results.}
    \label{fig: plot-hydro-random-im-vs-real-o4}
\end{figure}

Another intriguing aspect to consider is the comparison between the ratio of imaginary to real parts for both randomly generated modes and hydrodynamic modes. This comparison is depicted in Figure \ref{fig: plot-hydro-random-im-vs-real-o4}. The top row corresponds to $\tau_1$, while the bottom row represents $\tau_2$. Similarly, the left plots show LM results and the right plots display HM results. The ratio for hydrodynamic modes is consistently greater than that for random modes. For other scenarios, such as the 10th-order expansion, the results do not exhibit significant differences from those shown in Figure \ref{fig: plot-hydro-random-im-vs-real-o4}.

\begin{figure}
    \centering
    \includegraphics[valign =t, width=0.48\textwidth]{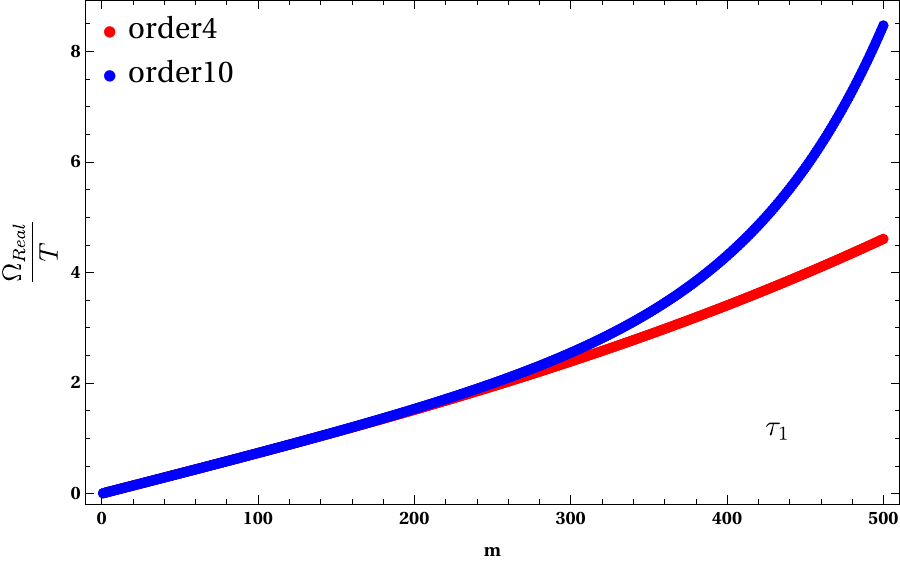} 
    \hspace{0.4cm}
     \includegraphics[valign =t, width=0.48\textwidth]{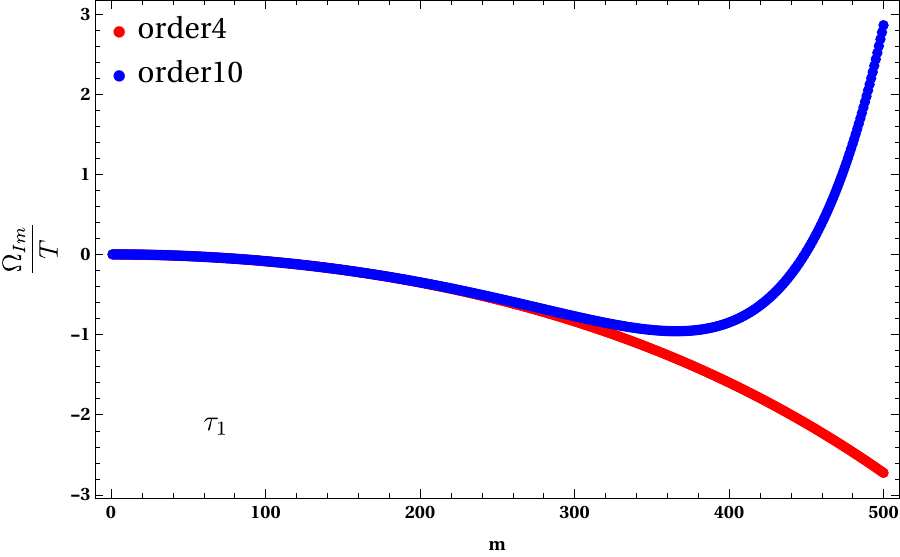}\\
     \vspace{0.5cm}
     \includegraphics[valign =t, width=0.48\textwidth]{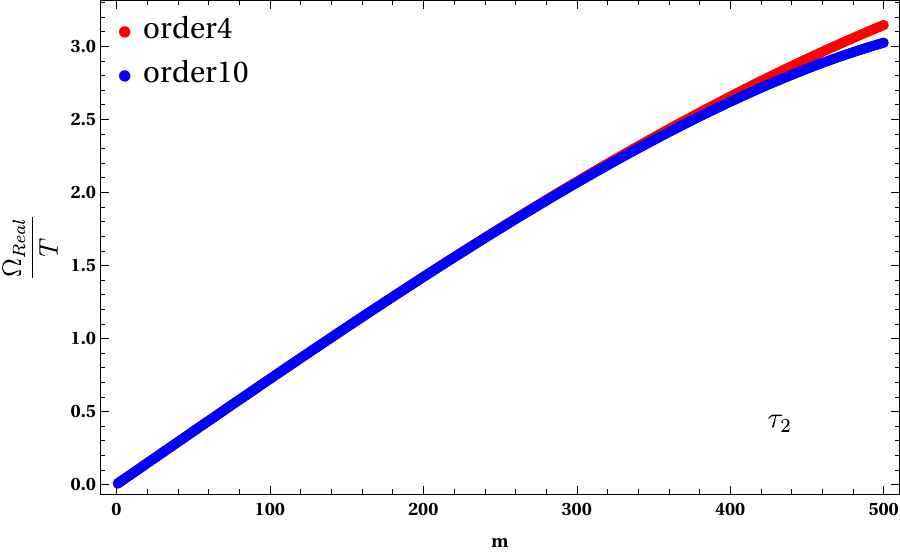} 
    \hspace{0.4cm}
     \includegraphics[valign =t, width=0.48\textwidth]{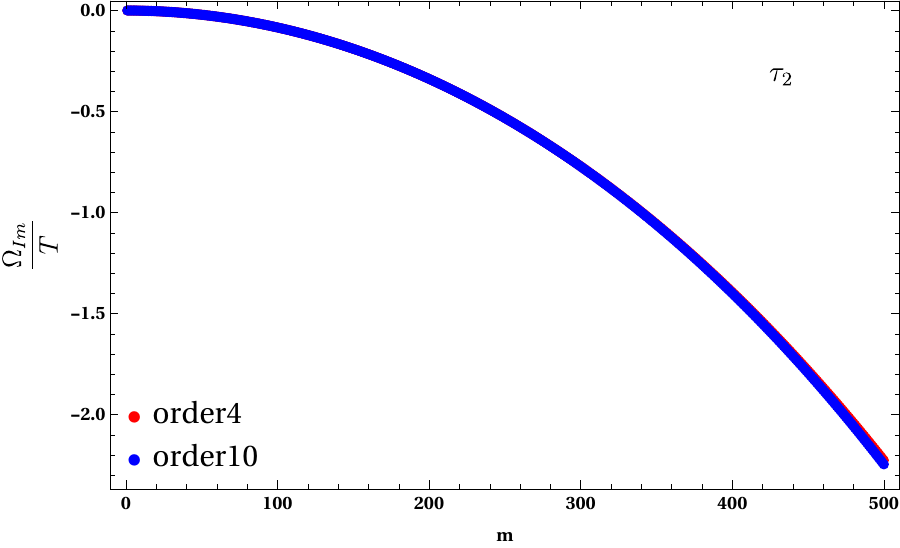}
    \caption{The evolution of the real and imaginary parts of the hydrodynamic modes in terms of momentum bin. The top row displays the results for $\tau_1$, and the bottom row shows the results for $\tau_2$. Different colors correspond to different orders of expansion as provided in \eqref{eq: mode-hydro-MIS} and \eqref{eq: hydro-MIS-o10}.}
    \label{fig: plot-o4-o10-hydro}
\end{figure}

As mentioned above, moving to the HM region would lose our predictive power, since the structure of solutions alters in the HM due to the presence of branch cuts \cite{Grozdanov:2018fic, Heydari:2024qzc}. This phenomenon is observed in Figure \ref{fig: plot-o4-o10-hydro}, where we illustrate the evolution of the real and imaginary parts of the hydrodynamic modes in both of the LM and HM regions for $\tau_1$ in the top row and $\tau_2$ in the bottom row. The red (blue) color traces the path of the 4th (10th) order series corresponding to \eqref{eq: mode-hydro-MIS} and \eqref{eq: hydro-MIS-o10}, respectively. In the LM region, the order of expansion does not significantly affect the results. However, in the HM region, the series order does impact the real and imaginary parts, particularly for $\tau_1$. This is because around $m \sim 300$, the solutions $\omega(k)$ transits from exhibiting two propagating and one dispersive mode to three propagating modes, a change attributed to the alteration in branch cuts. For $\tau_2$, this shift point moves approximately to $m \sim 1000$, where the modes undergo a similar change. Since the latter point is too distant to be displayed in Figure \ref{fig: plot-o4-o10-hydro}, the colors in the bottom plots are so close to each other.

Concluding this discussion should emphasize again that convergence and analyticity are somewhat distinct concepts.   In works such as \cite{gavmodes, rochamodes} it was argued that deterministic transport models generally have non-analytic cuts in the IR which do not impact the convergence of the Knudsen series.   In \cite{tails1, tails2} on the other hand perturbative fluctuations are shown to introduce both non-analyticity and lack of convergence cascading from the UV to the IR (``long-time tails'').
As argued in the introduction, soft non-analyticity is in principle indistinguishable from hydrodynamic modes when fluctuations are included, so our random coefficients include any soft modes in the $\ave{T_{\alpha \beta} T_{\gamma \mu}}$ correlator, which are ``by definition'' called hydrodynamic modes.  Our random coefficients ansatz is based on thermal fluctuations being treated as ``large'' w.r.t. the Kundsen series so the regime of validity is different from that discussed in \cite{tails1,tails2}. However, the correlation between convergence and analyticity as argued in \cite{Heydari:2024qzc}, suggests that provided convergence holds analyticity becomes a ``moot point'' since any non-hydrodynamic modes can be accounted for by fluctuating coefficients.  In this case, the effect of fluctuations on the distribution of imaginary to real poles can alter the impact of long-lived ``hydrodynamic'' modes considerably w.r.t. perturbative expectations.
\section{Conclusion}
As can be seen, the introduction of fluctuations consistently lowers the ratio of imaginary to real parts of the roots at all $\omega/T$, and makes the distribution of the roots ''scale-invariant'' with respect to $\omega/T$.  
Since, as far as we know, the type of series shown here has not been studied beyond numerics (exact mathematical results \cite{kac, berzin2022kacrice, Nicolaescu2022, Nguyen_2016} concern polynomials where all coefficients are of the same order) we are not aware for a mathematical justification for this, beyond a naive appeal to the Kac formula (where the fraction of real roots goes to zero but the probability of finding at least a real root goes to unitary as order increases).

The calculations here are certainly naive, only involving sound and shear modes rather than vorticity.  Nevertheless, if they capture the essential physics, the results are remarkable.  They provide evidence for the conjecture, made in \cite{Dore:2021xqq, functional, Zubarev} that hydrodynamic behavior does not necessarily go down with the number of constituents: Because the background fluctuates, and because the condition for long-range propagation is a delicate correlation of all coefficients in the dispersion relation, it might be that fluctuations allow asymptotic propagation to be ''selected out''.

There is certainly quite a lot of work to be done in this direction.  It will be interesting to see if the results are maintained as the order of the polynomial goes to infinity, connected to studied by other methods in \cite{grozfuture}.
Likewise, it will be interesting to see the probability structure of the analytical features (critical points, real vs imaginary roots) of fluctuating polynomials, and to relate this to the gap problem \cite{moore, gavmodes, rochamodes, roymodes}.
It would also be interesting to see whether dispersion relations with fluctuating coefficients can be used to quantitatively model ''collective'' systems with few degrees of freedom obtainable in more controlled laboratory conditions, such as \cite{smallhyd1,smallhyd2}.

The larger point here is that we do not know exactly how the hydrodynamic limit is approached when both the microscopic scale and the mean free path are varied \cite{ryb}.  
It is commonly assumed that even in the strongly coupled limit one gets closer to ideal hydrodynamics when one increases the number of degrees of freedom.  This is a reasonable assumption but there is no solid evidence for it and in fact the evidence in \cite{Nagle:2018nvi,smallhyd1,smallhyd2} pushes us to question it.  
The indications, discussed here,  that adding background fluctuations to ''hydrodynamic'' dispersion relations non-trivially changes long-distance propagaton could therefore profoundly affect our understanding of the onset of collectivity in small systems.

\section{Acknowledgements}
We sincerely appreciate Saso Grozdanov for reading the manuscript and clarifying some objective points. G.T.~acknowledges support from Bolsa de produtividade CNPQ 305731/2023-8, Bolsa de pesquisa FAPESP 2023/06278-2.

\appendix
\section{Random polynomials: A brief review}
A random function is defined as a probability measure on the space of functions that map from a set of parameters \cite{Nicolaescu2022}. By establishing a basis $(f_0(t), f_1(t), \cdots, f_N(t))$ (which does not necessarily have to be orthonormal) and selecting a random complex vector, such as $X= (x_0, x_1, \cdots, x_N)$, we can construct a random function $f(t)$ as follows
\begin{align}\label{eq: random}
    f: T \to \mathbb{R}, \qquad f(t) = \sum\limits_{i=0}^N x_i \, f_i(t).
\end{align}
This random function is characterized by the probability distribution that describes the likelihood of each event
\begin{align}
    &\mathbb{P} = p(x_0, x_1, \cdots, x_N) \, dx_0 \, dx_1 \cdots dx_N,\nonumber\\
    &x_i \in (\bar{x}_i, \,\bar{x}_i + dx_i),
\end{align}
where the probabilities are normalized to sum to one
\begin{align}
    \int\limits_{\text{All X}}\, p(x_0, x_1, \cdots, x_N) \, dx_0 \, dx_1 \cdots dx_N = 1.
\end{align}
The elements of the set $X$ can be generated from various random distributions, including uniform and Gaussian distributions. When using a uniform distribution, the elements are generated randomly and independently without any correlation. In contrast, other distributions, such as the Gaussian distribution, produce numbers that are correlated with each other.

The Fundamental Theorem of Algebra tells us that for any polynomial $f(t)$ of degree $n$ with coefficients in $\mathbb{C}$, there exist $n$ roots, counting multiplicity, such that $f(t) = 0$. This theorem leads to several intriguing questions:
\begin{enumerate}
\item  Given the distribution of coefficients $X$, what is the distribution of the roots of $f(t) = 0$? More specifically, what can we say about the distribution of the absolute values of the roots and their phases, assuming a certain distribution of coefficients?

\item How many real roots can we expect for $f(t) = 0$?

\item  What is the distribution of the real roots of $f(t) = 0$?
\end{enumerate}

These questions have sparked extensive research in the field of pure mathematics \cite{Nicolaescu2022, kac, berzin2022kacrice, Nguyen_2016}. For lower-degree polynomials, it is relatively straightforward to compute probabilities using closed-form solutions. However, the lack of general analytical solutions makes the task significantly more complex for higher-degree polynomials (specifically, those of degree $5$ or greater). 

The Kac-Rice formula provides a general solution to this problem. Consider a random function $f(t)$, as defined in \eqref{eq: random}, with a distribution given by
\begin{align}
    p_{f(t)}(X) dX = p_{f(t)}(x_0, \cdots, x_N)\,  dx_0 \cdots dx_N,
\end{align}
where the density $p_{f(t)}(X)$ is continuous at zero, and the coefficients $X$ are drawn from the set of real numbers. The expected number of real roots, denoted by $\hat{Z}(f, T)$, can be expressed as
\begin{align}\label{eq: expected-discrete}
    \hat{Z}(f, T) = \sum\limits_{k=0}^N \, k \, \mathbb{P}_R(k),
\end{align}
where $\mathbb{P}_R(k)$ is the probability of having $k$ real roots. The Kac-Rice formula offers an alternative expression
\begin{align}\label{eq: expected-continous}
   &\hat{Z}(f, T) = \int\limits_{T} dt \, C(t),\nonumber\\
   &C(t) = \textbf{E} (f'(t) \vert f(t)=0) \, p_{f(t)}(0),
\end{align}
where $\textbf{E} (f'(t) \vert f(t)=0)$ is the conditional expectation of the derivative $f'(t) = df(t)/dt$ given that $f(t) = 0$, and the integral is taken over the range of the random variable ``$t$''. The conditional expectation is defined as
\begin{align}\label{eq: conditional-expectation}
  &\textbf{E} (f'(t) \vert f(t)=0)  = \int\limits_{\mathbb{R}} dy \, |y| \, q_t(y),\nonumber\\
  &q_t(y) = \frac{p_t(0, y)}{p_{f(t)}(0)},\nonumber\\
  &p_{f(t)}(X) = \int\limits_{\mathbb{R}} dy \, p_t(X, y).
\end{align}
The \eqref{eq: expected-continous} and \eqref{eq: conditional-expectation} can be seen as the continuous analogs to the discrete formulation in \eqref{eq: expected-discrete}.

If we assume a Gaussian distribution for $f(t)$, the analysis becomes significantly simpler. The random vector $(f'(t), f(t)) \in \mathbb{R}^2$ follows a Gaussian distribution, which is characterized by its covariance matrix
\begin{align}
    C_t = \begin{pmatrix}
    a_t & b_t \\
    b_t & c_t
    \end{pmatrix},
\end{align}
where
\begin{align}
    a_t &= \textbf{E}(f'(t)^2) = \frac{\partial^2 K(s, t)}{\partial s \, \partial t}\Big|_{s=t}, \nonumber\\
    b_t &= \textbf{E}(f(t) f'(t)) = \frac{\partial K(s, t)}{\partial t}\Big|_{s=t}, \nonumber\\
    c_t &= \textbf{E}(f(t)^2) = K(t, t),
\end{align}
and $\textbf{E}(.)$ denotes the expectation of the term within the parentheses. Here, $K(s, t) = \textbf{E}(f(s) f(t))$ is the covariance kernel. Defining $\vec{F} \equiv (y, x) = (f'(t), f(t))$, the joint distribution is given by
\begin{align}
    \Gamma_{C_{t}} \, dx \, dy = \frac{1}{2\pi \sqrt{\Delta_t}} e^{-\frac{\vec{F}^T \cdot C_t \cdot \vec{F}}{2 \Delta_t}}\, dx \, dy,
\end{align}
with $\Delta_t = \det(C_t) = a_t c_t - b_t^2$. Substituting this joint distribution into \eqref{eq: expected-continous} and performing some calculations, we obtain \cite{Nicolaescu2022}
\begin{align}\label{eq: expected-Kac-Rice}
    \hat{Z}(f, T) = \frac{1}{\pi}\int\limits_{T} dt \, \rho_t,
\end{align}
where
\begin{align}
    \rho_t = \frac{\sqrt{\Delta_t}}{a_t} = \sqrt{\frac{\partial^2 \ln{K(s, t)}}{\partial s \, \partial t}\Big|_{s=t}}.
\end{align}
For example, consider the degree $N$ polynomial
\begin{align}
    f_N(t) = \sum\limits_{k=0}^N x_k t^k,
\end{align}
with a uniform distribution such that $\langle x_i x_j \rangle = \delta_{ij}$. The covariance kernel for this polynomial is
\begin{align}
    K_N(s, t) = \textbf{E}(f(s) f(t)) = \frac{1 - (st)^{N+1}}{1 - st},
\end{align}
and \eqref{eq: expected-Kac-Rice} yields
\begin{align}\label{eq: expected-uniform}
    \hat{Z}(f_N, T) = \frac{4}{\pi}\int\limits_{1}^\infty dt \, \sqrt{F_N(t)},
\end{align}
where
\begin{align}
    F_N(t) = \frac{1}{(1-t^2)^2} - \frac{(N+1)^2 t^{2N}}{(1-t^{2N+2})^2}.
\end{align}
For $N=2$, this gives $\hat{Z}(f_2) \simeq 1.279$, which is close to $1.256$. For $N=3$, it gives $\hat{Z}(f_3) \simeq 1.492$, which agrees well with numerical computations. For higher values of $N$, \eqref{eq: expected-uniform} provides results close to those obtained numerically. It has been argued that the asymptotic behavior of the expected number of real roots in a random real function with a uniform distribution as $N$ becomes very large is given by \cite{Nguyen_2016}
\begin{align}\label{eq: estimation}
    \hat{Z}(f_N)\Big|_{N \to \infty} \simeq \frac{2 \ln{(eN)}}{\pi}.
\end{align}
It is important to note that \eqref{eq: expected-Kac-Rice} can be applied to many random functions with different distributions, provided that the covariance kernel can be derived analytically. However, the Kac-Rice formula does not provide information about the distribution of real roots or the distribution of roots in general.
\medskip

\bibliographystyle{fullsort}
\bibliography{refs}
\end{document}